\documentstyle[preprint,prb,aps]{revtex}

\begin{document}
\draft
\preprint{IUCM96-006}

\title{Hartree-Fock Theory of Skyrmions in Quantum Hall Ferromagnets}

\author{H.A. Fertig$^{1}$, Luis Brey$^{2}$, R. C\^{o}t\'{e}$^{3}$,
A.H. MacDonald$^{4}$, A. Karlhede$^{5}$ and S.L. Sondhi$^{6}$.}
\address{$^{1}$ Department of Physics and Astronomy, University of
Kentucky, Lexington KY 40506-0055}
\address{$^{2}$ Instituto de Ciencia de Materiales (CSIC),
Universidad Aut\'{o}noma C-12, 28049 Madrid, Spain}
\address{$^{3}$ D\'{e}partement de Physique et Centre de
Recherches en\\
Physique du Solide, Universit\'{e} de Sherbrooke, Sherbrooke,\\
Qu\'{e}bec, Canada J1K-2R1}
\address{$^{4}$ Department of Physics, Indiana University,
 Bloomington IN 47405}
\address{$^{5}$ Department of Physics, Stockholm University,
Box 6730, S-11385 Stockholm, Sweden}
\address{$^{6}$ Department of Physics, Princeton University,
Princeton NJ 08544}

\date{\today}
\maketitle

\begin{abstract}

We report on a study of the charged-skyrmion or spin-texture
excitations which occur in quantum Hall ferromagnets near odd
Landau level filling factors.
Particle-hole symmetry is used to relate the spin-quantum numbers of
charged particle and hole excitations and neutral particle-hole pair
excitations.  Hartree-Fock theory is used to provide quantitative estimates of
the energies of these excitations and their dependence on
Zeeman coupling strength, Landau level quantum numbers,
and the thicknesses  of the two-dimensional electron layers.
For the case of $\nu$ near three we suggest
the possibility of first order phase transitions with increasing
Zeeman coupling strength from a many skyrmion state
to one with many maximally spin-polarized quasiparticles.

\end{abstract}


\section{Introduction}

In the limit of strong magnetic fields where all electrons are confined to
a single orbital Landau level, interacting two-dimensional electron
systems (2DES's) exhibit a rich variety of unusual properties. The quantum
Hall effect, which occurs when the chemical potential of the 2DES has a
discontinuity at a density $n^{*}$ which depends  on magnetic field
strength\cite{leshouches}, is a prominent example.  In the quantum Hall
effect the incompressible  groundstate at density $n^{*}$ can be a strong
ferromagnet, {\it i.e.}, its total spin quantum number $S$ can equal $ N/2 $
so that electronic spins are completely aligned by infinitesimal
Zeeman coupling.  Surprisingly, in light of
the strong magnetic fields, physics associated with {\it spontaneous}
magnetization  in these systems is experimentally accessible because the
Zeeman  coupling is typically quite weak compared to other characteristic
energy scales and can even be tuned to zero, for example by the
application of hydrostatic pressure to the host semiconductor.  Recently
there  has been considerable interest in charged excitations  of the
incompressible groundstate at $\nu =1$ which has total spin quantum
number $S = N/2$.  As first noticed in numerical exact diagonalization
calculations,\cite{ednumerical} and demonstrated in recent
experiments\cite{barrett,goldberg} the  spin-polarization in
these systems is strongly reduced away from $\nu=1$ where the groundstate
must incorporate the charged excitations of the $\nu=1$ state.
This behavior can be understood,\cite{sondhi,hf} quantitatively in the
limit of very weak Zeeman coupling, by identifying the elementary charged
excitations with the topological solitons (skyrmions) of the $O(3)$
non-linear sigma ($NL\sigma$) model in two spatial dimensions\cite{rajaraman}.
With an appropriate kinetic term\cite{fradkin}, the latter describes the
long wavelength $T=0$ dynamics of any Heisenberg ferromagnet in two dimensions.
Two features distinguish the quantum Hall case. First, the skyrmions carry
electrical charge as a consequence of their topological
charge,\cite{kane,sondhi,kramerasi} and hence
have a stable finite size for small but non-zero
Zeeman coupling. Second, they are present in the groundstate near
(but not precisely at) $\nu = 1$, and as a
consequence have an obvious influence on observable properties.

In this article we discuss the elementary charged excitations of  quantum
Hall ferromagnets using a Hartree-Fock approximation
approach.\cite{hf,moon}  The Hartree-Fock  approximation can describe
charged excitations  which are sufficiently localized in space to
invalidate the gradient expansion that underlies the $NL\sigma$ model
description. A weakness of both the Hartree-Fock
and $NL\sigma$ model calculations, which we will
discuss further below,
is the failure to respect the quantization\cite{nayak,hcmskyrm,oaknin}
of total electronic spin. The article is organized as follows.
In Section II we  make some general
remarks on the implications of particle-hole symmetry for relationships
between the energies and  spin quantum numbers of positively and negatively
charged excitations of quantum Hall ferromagnets and for the  neutral
decoupled particle-hole pair excitations which are  important in activated
transport experiments.\cite{schmeller} The considerations in this section
do not depend on the Hartree-Fock approximation.  In Section III we
explain some formal aspects  of the Hartree-Fock approximation
calculations we perform
in order to explore the elementary charged excitations and how their
energetic ordering depends on Zeeman coupling strength.  Section IV
presents and discusses our numerical results. We comment on the importance
of the orbital Landau level index\cite{wu} and on the thickness of
two-dimensional electron layers.  We conclude with a brief
summary in Section V.  An appendix discusses the spectral transfer
in the presence of the skyrmion, analogous to that underlying midgap
states in solitons in one-dimensional systems.

We close this section with a remark on the terminology  used in this
paper.  In $NL\sigma$ models the energy of a ferromagnet is expressed
in terms of a function which specifies the direction of the
local ordered moment as a  function of the two-dimensional spatial
coordinate.  The energy functional in the ``pure'' $NL\sigma$ model
invoked in discussions of broken symmetry states contains only
a gradient term and the topological solitons of this model can be
determined exactly.\cite{rajaraman}  As the gradient term is
scale invariant in two dimensions, the energy of these solutions
is independent of their size. The $NL\sigma$ model appropriate
for quantum Hall ferromagnets with a small Zeeman coupling has two
additional terms which compete: a Zeeman coupling which favors small
skyrmions and a Coulomb interaction which favors
large skyrmions; together these determine the size and energy of the
skyrmion and also its precise profile which differs from those of the
pure $NL\sigma$ model solutions\cite{fn1}. In a microscopic quantum
treatment\cite{hcmskyrm,ednumerical} a spin-multiplet of
elementary charged excitations with total spin $S = N/2 - K $ exists for each
non-negative integer $K$. At large values of $K$, the correlations of
the quantum states are well described by the classical soliton
solutions; presumably, a treatment of the fluctuations about them,
along the lines of soliton quantization in other problems\cite{rajaraman},
would reproduce the exact states. However, frequently the lowest energy charged
excitations occur at  relatively small integer values of $K$  where both
the symmetry restoring quantum fluctuations and the neglected higher gradient
terms are large and the field theoretic description is no longer accurate.
While, strictly speaking, one might wish to reserve the term ``skyrmion''
for excitations that are well described by the $NL\sigma$ model solitons,
in this paper we take the liberty of referring to all elementary charged
excitations of quantum Hall ferromagnets as skyrmions.

\section{Particle-Hole Symmetry}

When the spin-degree of freedom is included, the particle-hole symmetry of
the Hamiltonian\cite{leshouches} for interacting electrons in the lowest
Landau level relates electronic states at Landau level filling factors
$\nu$ and $2 - \nu$. Here $\nu \equiv N/N_{\phi}$ where $N$ is the number
of electrons and $N_{\phi} = A B / \Phi_{0} \equiv A/ (2 \pi \ell^{2}) $ is
the  orbital degeneracy of the Landau level.  ($B$ is the magnetic  field
strength, $\Phi_{0} = hc/e $ is the electronic magnetic  flux quantum, and
$\ell$ is the magnetic length.) The particle-hole symmetry of this system
occurs because, apart from the constant quantized kinetic energy which is
conventionally chosen as the  zero of energy, the Hamiltonian contains
only the  term describing interactions of electrons within a degenerate
Landau level.  In the occupation number representation, many-particle
states can be specified either by the set of single-particle states within
the Landau level which are occupied or by specifying the  set which are
empty, {i.e.} by specifying the states occupied by holes in the Landau
levels.  It is convenient to combine the particle-hole  transformation
with a spin-reversal so that the creation operators for spin-up particles
are mapped to annihilation operators for spin-down holes and creation
operators for spin-down particles are mapped to annihilation operators
for spin-up holes.  Under this mapping
\begin{eqnarray}
N_{\uparrow} &\to& N_{\phi} - N_{\downarrow} \equiv
N'_{\uparrow}\nonumber\\
 N_{\downarrow} &\to& N_{\phi} - N_{\uparrow} \equiv N'_{\downarrow}.
\label{eq:phmapping}
\end{eqnarray}
It follows that $N = N_{\uparrow} + N_{\downarrow}
\to N' = 2 N_{\phi} -N$ and $S_{z} = (N_{\uparrow} - N_{\downarrow})/2
\to S'_{z} = S_{z}$ where $S_{z}$ is the $\hat{z}$  component of the total
spin.  The Hamiltonian $H$ changes by a constant:\cite{phsym}
\begin{equation}
H  = H' + 2 (N -N_{\phi})  \epsilon_{0} = H' + 2 (N_{\phi} - N')
\epsilon_{0}
\label{eq:hpmapham}
\end{equation}
where $H'$ is normal ordered in terms of hole creation and annihilation
operators and is identical to $H$ except for the replacement of electron
operators by hole operators, and $\epsilon_{0}$ is the energy per electron
in the groundstate at $\nu=1$ which is readily calculated for any
specified model electron-electron interaction.   For 2D electron systems
with a uniform neutralizing positive background charge
\begin{equation}
\epsilon_{0} = -{1 \over 2} \int \frac{d^{2} \vec{q}}{(2 \pi)^{2}}
\exp{( - |q|^{2} \ell^{2} /2)} v(q)
\label{eq:epsilon_{0}}
\end{equation}
where $v(q)$ is the Fourier transform of the effective electron-electron
interaction.

In the following sections we will evaluate the energy change in the
system when a single electron is removed from the system at a fixed
magnetic field.  As mentioned above and shown in
Reference\onlinecite{hcmskyrm}, the
set of elementary charged excitations at $N=N_{\phi}-1$
is composed of a single spin-multiplet with
$S=N/2 - K$ for each non-negative integer $K$.
With finite Zeeman coupling, the lowest energy state in
each multiplet has $S_{z} = S = N/2 -K = N_{\phi}/2 - (K+1/2)$ and  the
energy of this state relative to the $\nu =1$ groundstate of the quantum
Hall ferromagnet may be written as
\begin{equation}
\epsilon_{K}^{-} = U_{K} + g^{*} \mu_{B} B (K + 1/2) .
\label{eq:epsilonkm}
\end{equation}
Here $g^{*}$ is the g-factor of the host semiconductor,
and $U_{K}$ may be interpreted as the interaction contribution
to the internal energy of the quasihole.
It follows from
the particle-hole transformation (Eq.~(\ref{eq:phmapping})) that the
corresponding elementary charged excitations at
$N= N_{\phi}+1$ have $S_{z} = S = N'/2-K = N_{\phi}/2 - (K+1/2)$ and energy
relative to the $N=N_{\phi}$ groundstate given by
\begin{equation}
\epsilon_{K}^{+} = U_{K} + 2 \epsilon_{0} + g^{*} \mu_{B} B (K+1/2).
\label{eq;epsilonkp}
\end{equation}
For example, the state with $K=0$ is the single-Slater  determinant
elementary excitation which appears in the standard Hartree-Fock approximation.
Note that $\epsilon_{K}^{-}$ and $\epsilon_{K}^{+}$ differ by  a constant
so that their minima always occur at the same value of $K$.  For
two-dimensional electrons in the lowest Landau level with purely
Coulombic interactions, $U_{K=0} =
(e^{2}/\epsilon \ell) (\pi/2)^{1/2}$, and $\epsilon_{0}= - (e^{2}/\epsilon
\ell) (\pi/8)^{1/2}$ so that
$\epsilon_{K=0}^{-} = U_{K=0} + g^* \mu_B B /2 $
and $\epsilon_{K=0}^{+} =  g^{*} \mu_{B} B/2 $.
It follows\cite{sondhi} from the $NL\sigma$  model that $U_{K \to
\infty} = (3/4) U_{K=0}$ so that large
$K$ states will have lower energy if the Zeeman coupling is sufficiently
weak.  At intermediate  values of $K$ numerical calculations of one sort
or another are  necessary to estimate $U_{K}$.  In the following section
we  use a generalized Hartree-Fock approximation to obtain realistic
estimates of $U_{K}$ including effects of the finite  thickness of the
quasi two-dimensional electron layers and  to discuss the influence of the
Landau level index for which  the quantum Hall ferromagnet occurs.

The value of $K$ at which $\epsilon_{K}^{-}$ and
$\epsilon_{K}^{+}$ are minimized has been determined experimentally  both
by measuring the groundstate spin-polarization\cite{barrett,goldberg}
near $\nu =1$ and by measuring\cite{schmeller} the dependence on Zeeman
coupling  strength of energy required for the creation of free-particle
hole pairs at $\nu = 1$.  For $\nu$ close to one, interactions between
elementary charged excitations can be neglected\cite{skyrmlat} so  that
the partial spin polarization is given for $\nu < 1$ by
\begin{equation}
\xi \equiv \frac{N_{\uparrow} -
N_{\downarrow}}{N_{\uparrow}+N_{\downarrow}} = \frac{2 S_{z}}{N} \approx
1 - 2K(1 - \nu) / \nu
\label{eq:xim}
\end{equation}
and for $\nu > 1$ by
\begin{equation}
\xi \equiv \frac{2 S_{z}}{N} \approx  1 - 2(K+1) (\nu -1 ) / \nu .
\label{eq:xip}
\end{equation}
Note that $d \xi / d \nu =2K$ just below $\nu =1$ and
$- 2 (K+1) $ just above $\nu =1$.
The polarization $\xi$ may be measured in
experiments\cite{barrett,goldberg} that
are sensitive to the spin magnetization of the system.
Finally, the activation gap $\Delta$  measured
in transport studies\cite{schmeller}
is the energy to make an unbound particle-hole pair:
\begin{equation}
\Delta = \epsilon_{K}^{+} + \epsilon_{K}^{-}
       = 2 U_{K} + 2\epsilon_{0} + g^{*} \mu_{B} B (2 K +1).
\label{eq:delta}
\end{equation}
The derivative of this energy with respect to the strength ($ g^{*}
\mu_{B} B $) of the Zeeman coupling term is $2 K +1$. As we discuss at
greater length below, experiments\cite{barrett,goldberg,schmeller}
performed at  typical fields in GaAs
2DES's are all consistent with $K = 3$,
in good agreement with calculations presented in subsequent sections.

\section{Quasiparticle Energies and Excitation Gaps}

In this section some formal aspects of the Hartree-Fock theory for the
the charged excitations at $\nu < 1$ are briefly discussed.  Further
technical details are provided in a previous paper on this
topic \cite{hf} and in the appendix.  (The excitations at
$\nu > 1$ may be obtained from these by particle-hole  symmetry as
explained above.)  Because of the symmetry of the  skyrmion charged
excitations, it is convenient to work in the symmetric gauge where
single-particle wavefunctions in the lowest  Landau level (LLL) have the
simple form
\begin{equation}
\phi_{m}(z) = \frac{z^{m} \exp ( - |z|^{2}/ 4 \ell^{2})}{(2^{m+1} \pi
\ell^{2} m!)^{1/2}}.
\label{eq:pswfs}
\end{equation}
Here $m = 0 ,1 , \ldots, $ is the angular momentum, and $z = x + iy$ expresses
the 2D coordinate as a complex number.  Note that states with  larger $m$
are localized further and further from the origin. We consider
single-Slater determinant states of the form
\begin{equation}
|\Psi\rangle = \prod_{m=0}^{\infty} (u_{m} a_{m}^{\dagger} +v_{m} b_{m +
1}^{\dagger}) |0>
\label{eq:slatdet}
\end{equation}
where $|0\rangle$ is the particle vacuum and
$|u_{m}|^{2}+|v_{m}|^{2} = 1$ so that the wavefunction is normalized.
Here $a_{m}^{\dagger}$ creates a down-spin electron and
$b_{m}^{\dagger}$ creates an up-spin electron in the $mth$ angular
momentum state.

The form of these wavefunctions is essentially dictated by the
symmetry of the classical skyrmion solutions which are invariant
under the action of $L_z \pm S_z$ for the skyrmion (antiskyrmion).
This plus the requirement of LLL occupancy uniquely picks
Eq.~(\ref{eq:slatdet}) as the appropriate Hartree-Fock wavefunctions.
It is also easy to demonstrate using Eq.~(\ref{eq:slatdet}) that the
expectation value of the total spin-operator in this state describes a
spin-texture with unit topological charge provided that
$u_{m}$ varies slowly with $m$ from
$u_{m=0}=1$ to $u_{m \to \infty}=0$. ($v_{m=0}=0$ and $v_{m \to \infty}=1$).

This wavefunction is a generalization of the single-Slater  determinant
wavefunction proposed\cite{moon} by Moon {\it et. al.} for which the
expectation value of the vector spin operator gives a spin-texture
identical with that of the pure $NL\sigma$ model skyrmion. The additional
variational freedom in the wavefunction of  Eq.~(\ref{eq:slatdet}) allows
deviations from the pure $NL\sigma$ model so that microscopically localized
charged excitations may be optimized in a way which depends on the  details of the
2DES being considered.  Far from the origin, this state is locally
identical to the  ferromagnetic groundstates and all spins are aligned
with the  Zeeman magnetic field which is assumed to point in the `up'
direction. Near the origin the projection of the total spin along the
field  direction becomes negative.  It is easy to establish that the total
decrease in electron charge near the origin compared to the ferromagnetic
groundstate ($\prod_{m=0}^{\infty} b_{m}^{\dagger} | 0 \rangle$)
corresponds to one electron.  The total number of reversed spins in this
wavefunction is
\begin{equation}
K = \sum_{m=0}^{\infty} |u_{m}|^{2}.
\label{eq:khf}
\end{equation}
The $K=0$ Hartree-Fock hole excitation is obtained by choosing
$u_{m}=0$ for all $m$.  Note that the quantization of the number of
reversed spins is not captured by these generalized Hartree-Fock
variational wavefunctions.  The quantization is obviously of some
importance in the usual experimental circumstance since $K$ is small.
In this case better variational wavefunctions could be
obtained, in principle, by projecting\cite{nayak,hcmskyrm} the Hartree-Fock
wavefunctions onto states with definite numbers of reversed spins.
Here we account for quantization when necessary simply by restricting
ourselves to solutions where $K$ (the {\it mean} number of reversed
spins) is an integer\cite{ring}.

We find optimized skyrmion wavefunctions by minimizing
$\langle \Psi | H | \Psi \rangle$, where the Hamiltonian includes
electron-electron interactions and Zeeman coupling.
This procedure gives rise to
a set of self-consistent equations listed in Ref.~\onlinecite{hf}
which need to be solved numerically.
Some typical results for the spin and charge densities
relative to the groundstate are illustrated in
Fig.~\ref{fig:density}, for skyrmions
in the lowest Landau level in a sample of
vanishingly small thickness, for several
values of\cite{comhf}~ $\tilde{g}$,
where $\tilde g \equiv g^{*} \mu_{B} B$
is the Zeeman coupling strength\cite{comhf2}.
The total energy of this state,
relative to the energy of the ferromagnetic groundstate, can be separated
into interaction energy and Zeeman energy contributions as in
Eq.~(\ref{eq:epsilonkm}):
\begin{equation}
\epsilon^{-} (K(\tilde g)) = U (K (\tilde g )) + \tilde g [K+{1 \over 2}].
\label{eq:epsilonkhf}
\end{equation}
For each value of $\tilde g$ the self-consistent Hartree-Fock equations
determine the value of $K$ which minimizes $\epsilon$  and, given $K$, the
shape of the skyrmion state (specified by the $u_{m}$ values) which
minimizes the interaction energy. For pedagogical purposes and for
comparison with the field theoretical approach it is  useful to
consider the interaction energy as a function of $K$
rather than $\tilde g$.  $U(K)$, which is just the Legendre transform of
$\epsilon^-(K)$, can be obtained from the $K(\tilde g)$
and $\epsilon^-(K(\tilde g))$ produced by the  self-consistent
Hartree-Fock calculations using Eq.~(\ref{eq:epsilonkhf}). Formally
minimizing Eq.~(\ref{eq:epsilonkhf}) with respect to $K$ we find that
the optimal value of $K$ for a given $\tilde g$ is determined by
\begin{equation}
\tilde g = - \frac{d U(K)}{d K}.
\label{eq:selfcons}
\end{equation}
Thus the  global relationship
between the equilibrium value of $K$ and  the Zeeman coupling strength is
conveniently summarized in a plot of ${{dU} \over {dK}}$
versus $K$.  We
will use such plots in the next section to discuss the importance of the
finite thickness of two-dimensional electron layers  and of the Landau
level index of the ferromagnetic state in  determining $K (\tilde g)$.
The naive expectation is that the lowest energy skyrmion state should
monotonically shrink as $\tilde g$  increases; we see from
Eq.~(\ref{eq:selfcons}) that this is  possible only if $d^{2} U(K)/d^{2} K
$ is positive definite which is the usual requirement of convexity needed
to obtain a continuous Legendre transform. The maximum value of
$ - d U(K) /d K$ gives the maximum value  of $\tilde g$ at which Hartree-Fock
skyrmion states with $K\ne 0$  occur.

\section{Hartree-Fock Theory Numerical Results}
\subsection{Lowest Landau Level: n=0}

In Fig.~\ref{fig:one} and Fig.~\ref{fig:two} we show results for
$\epsilon(K(\tilde g))$, $\Delta(K(\tilde g))$ and
$K(\tilde g)$ obtained by solving the Hartree-Fock equations at a series
of $\tilde g$ values for the case of a strictly  2D electron system and
for the case of a quasi 2D electron system with the finite layer thickness
modeled to approximate the typical experimental situation.
Finite thickness can be introduced into the calculation
by assuming all the electrons are in the some confined
state $\chi(z)$ of either the inversion layer or quantum well
in which they reside.  The effective two-dimensional
electron-electron interaction may then be written
in the form
$$
v(\vec{r}) = \int \frac{d^2k}{(2\pi)^{2}}
e^{i\vec{k}\cdot \vec{r} } \int dz_{1} dz_{2}
|\chi(z_{1})|^{2} |\chi(z_{2})|^{2} \frac{2\pi e^{2}}{k}
e^{-k|z_{1}-z_{2}|},
$$
which is softer than the Coulomb interaction
at short distances, but at large distances approaches
the $1/r$ form.  The precise wavefunction
chosen for our calculation is the Fang-Howard\cite{fang}
form $\chi(z)=\left[{1}{2W^{3}}\right]^{-1/2}
z e^{-z/2W}$, where $W$ may be understood as a
measure of the thickness of the two-dimensional layer.
Finite thickness corrections tend to reduce the
energy scales of this system; for example,
$\epsilon_{0} = - 0.6267$ for the strictly 2D model and
$\epsilon_{0} = - 0.3963$ for the quasi 2D finite-thickness model
for $W=0.45\ell$.
However, we have found that the optimal values for the total
number of flipped spins in the skyrmion states are not dramatically
affected by finite thickness.

In Fig.~\ref{fig:one}, we see that both $\epsilon$ and
$\Delta$ rise rapidly toward their $K=0$ values as $\tilde g/
(e^{2}/\epsilon \ell)$ increases.
The rapid increase in the energy with increasing $\tilde g$
is associated with the rapid  shrinking of the optimal size of the
skyrmion charged excitation, as seen in Fig.~\ref{fig:two}.
For typical experimental systems\cite{barrett,goldberg,schmeller}
$\tilde{g}$ $\approx 0.015-0.020e^{2}/\epsilon\ell$ and the
measured number of flipped spins $K \approx 3$,
in good agreement with the present numerical results.
The small Zeeman coupling strength required to reduce
skyrmions to microscopic size is at first sight surprising and
reflects the relatively weak $K$ dependence of $U(K)$.
As explained in the
previous section, $U(K=0) - U(K=\infty) = (\pi/32)^{1/2} (e^{2}/\ell)$
for the Coulomb interaction model.
For large $K$ the principal correction to the NL$\sigma$ model
$K \to \infty$ value for $U(K)$  is the Coulomb self-interaction
of the excess charge.  Assuming that the spin and charge structure
of the skyrmion have the same size this energy is $\sim (e^{2}/\ell)
K^{-1/2}$.
{}From Eq.(~\ref{eq:selfcons}) the optimal
value of $K$ implied by this approximation is $ \propto  \tilde g^{-2/3}$
while the change in $\epsilon$ is proportional  to $\tilde g^{1/3}$.
A more careful  analysis\cite{sondhi} by Sondhi {\it et
al.} replaces $\tilde g$ in these  estimates by $\tilde g |ln (\tilde
g)|$.   The $\tilde g^{1/3}$  behavior at small $\tilde g$ means that even
weak Zeeman coupling compared to the characteristic interaction energy is
sufficient to eliminate most of the energy difference between the lowest
energy charged excitation and the $K=0$ Hartree-Fock quasiparticle energy.

{}As explained in the previous section
we can extract results for $U(K)$, the  internal
energy of a skyrmion hole as a function of the  number of reversed spins,
from Fig.~\ref{fig:one} and Fig.~\ref{fig:two}.  These results are shown in
Fig.~\ref{fig:three}.  Note that $d^{2} U(K) /d K^{2}$ is positive definite.
The property that $K(\tilde g)$ is not
markedly altered by finite-width corrections to the effective
interaction is associated with the weak $K$ dependence of the
difference between the two curves in this figure.
The results shown here hint at a practical technical
difficulty with these Hartree-Fock calculations.  In practice we are
forced to truncate the set of single-particle angular momenta we include
in our wavefunctions for both the ferromagnetic groundstate and for the
charged excitation at a finite value of $m = M_{\rm max}$.  Since the
single-particle orbital with angular momentum $m$ is localized near a ring
with  radius $\ell (2 (m+1))^{1/2}) $ this corresponds to working with  a
finite size electron disk of radius $R \approx \ell (2 M_{\rm max})^{1/2}$.
The energy difference between the ferromagnetic groundstate and the state
with the charged excitation will be given accurately by our calculation if
the tail of the disturbance associated with the charged excitation does
not extend to the edge of our system.  As the Zeeman coupling weakens and
the skyrmions get larger, finite-size effects become increasingly
important.  It is for this reason that  the $U(K)$ results shown in
Fig.~\ref{fig:three} do not  accurately approach the analytically known
$K \to \infty$ asymptote which for the case of strictly 2D interactions
has the value  $(9 \pi /32)^{1/2} (e^{2}/\epsilon \ell)
\approx 0.94 (e^{2}/\epsilon \ell)$. (It is possible however, to get an
accurate value for this limit by regularizing the $\tilde g=0$ skyrmion
by placing it on a sphere. This is sketched in the appendix.) The smaller
the value of $\tilde g$ the larger the value of $M_{\rm max}$ required to
get accurate
results.  The convergence properties of our calculations with respect to
$M_{\rm max}$ are illustrated in Fig.~\ref{fig:four} where
$K$ is plotted as a
function of $ (\tilde g |\log (\tilde g)|)^{-2/3}$. {}From the asymptotic
analysis\cite{sondhi} we know that for
$M_{\rm max} \to \infty$ this curve will follow a roughly straight line for
large
$K$.  For each finite $M_{\rm max}$ the numerical results bend away  from this
line, underestimating the number of reversed spins in the lowest energy
charged excitation.  From Fig.{~\ref{fig:four} we estimate that
calculations performed with
$M_{\rm max} = 120$ are accurate for $\tilde g > 0.01 (e^{2}/ \ell)$,
calculations with $M_{\rm max} = 480$ are accurate for
$\tilde g > 0.005 (e^{2}/\ell)$ and that calculations with
$M_{\rm max} = 1920$ are accurate for $\tilde g > 0.002  (e^{2}/ \ell)$.
The number of angular momentum states
required to obtain good convergence for the total number
of flipped spins $K$ is much larger than $K$, despite the
fact that the $u_{m}$ and $v_{m}$ approach their
groundstate values relatively rapidly for $m > K$.
As expected from the variational nature of the Hartree-Fock calculations,
the energies presented in Fig.~\ref{fig:one}
and in Ref.~\onlinecite{hf},
approach their asymptotic values
much more rapidly with increasing $M_{\rm max}$ than
estimates of the optimal value of $K$.
Finally, we note that, except where the contrary is explicitly stated,
results reported here and in Ref.~\onlinecite{hf}
were obtained with $M_{\rm max}=120$.  In particular this means that the
values of $K$ presented in Fig. 2 below
$\tilde g = 0.01 (e^{2}/ \ell)$ are slightly
underestimated; however, we have found no significant
errors in the energies of Fig. 1 for
$\tilde g > 0.002 (e^{2}/ \ell)$.

In Fig.~\ref{fig:three} it is interesting to note
that $U(K)$ approaches $K=0$ with a finite slope.
This is seen most clearly in Fig.~\ref{fig:two}
which, using Eq.(~\ref{eq:selfcons}),
can be regarded as a plot of $K$ {\it vs.} $ - dU(K)/dK$.
The maximum slope of $U(K)$ occurs at $K=0$ and since
$|dU/dK|$ increases monotonically with decreasing $K$ this slope
specifies the largest value of $\tilde g$ for which a
$K \ne 0$ solution of the Hartree-Fock equations occurs.
The property that the Hartree-Fock skyrmion
decreases
continuously to zero size with increasing $\tilde g$
contrasts with properties which would follow from other
forms for $U(K)$.  For example if $U(K)$ approached its
$K=0$ value from below quadratically, the maximum value of
$-dU(K)/d(K)$ would occur at a finite value of $K$ and
the skyrmion would suddenly collapse to zero size
once $\tilde g$ exceeded this value.  If $U(K)$ approached
its $K=0$ value from below as $K^{s}$ with $s < 1$,
solutions with finite $K$ would exist at arbitrarily large
$\tilde g$.  Finally if $U(K)$ was an increasing function
of $K$ at small $K$, reaching a maximum at $K=K^{*}$,
solutions with $K < K^{*}$ would not exist at any
$\tilde g$.  We will see below that this is precisely the
situation which often occurs when the quantum Hall
ferromagnet groundstate has electrons in $n \ne 0$
Landau levels polarized.

As remarked in the introduction, it is important to
note that our Hartree-Fock approach does not
respect the separate quantization of spin and orbital angular
momentum in our system, and instead permits only the
quantization of the difference of these two quantities.
In the Hartree-Fock approximation, as in classical field
theories, skyrmions appear as broken symmetry states of the
Hamiltonian.  There is a similarity here to the standard BCS treatment
of superconductivity, in which particle number is not a good
quantum number in the mean-field groundstate\cite{moon,nayak}.
However, while in the BCS problem there really is a broken
symmetry in the thermodynamic limit, in our problem the
skyrmion states only break the symmetry over a finite volume.
Hence quantum fluctuations around the mean-field state will,
if treated exactly, restore the individual quantization of spin
and angular momentum. It is therefore important to address the
extent to which these fluctuations influence the results in
Fig.~\ref{fig:one}.  In an exact calculation Fig.~\ref{fig:one} would
take the form of a series of straight line segments, since the
quantum number $K$ can only take on integer values\cite{hcmskyrm}.
We expect the Hartree-Fock approach to yield a smooth interpolation
of the exact results which is most accurate when its $K$ value is
integral.  Comparison with exact finite size calculations~\cite{he,juanjo}
confirm this expectation. (We note the fortunate feature that the
Hartree-Fock treatment is exact both at vanishing $\tilde g$, where the
infinite skyrmion {\em is} a classical object up to trivial global
rotations and at large $\tilde g$ where it describes the fully polarized
quasihole.)
\bigskip

\subsection{First Landau Level: $n=1$}

\bigskip
Quantum Hall ferromagnets occur at $\nu=1$ and also
at larger odd integral filling factors.
In the Hartree-Fock approximation, the groundstate for
$\nu =2 n +1$ has all orbitals of both spins
with Landau level index less than $n$ occupied and only majority spin orbitals
occupied in the $n$-th Landau level.  If Landau level
mixing is ignored, the fully occupied Landau levels play
no role and the theory of quantum Hall ferromagnets is altered only
by the change in the index of the Landau level onto which
the electronic Hilbert space is projected.
The Hartree-Fock single-Slater-determinant state
is an exact eigenstate of the Hamiltonian just as in the
$\nu =1$ case and is expected to
be the groundstate, at least if $n$ is not too large.
The change of Landau level index may be accounted for
without approximation simply by
including an additional form factor
correction to the effective electron-electron interaction.
The Fourier transform of the interaction for electrons projected into
the $n$th Landau level is\cite{haldane}
$v_{n}(q) =  v(q) [L_{n}(\frac {q^{2}} {2})]^2$, where
$L_{n}$ is the $n$th Laguerre polynomial, and $v(q)$ is the
unprojected interaction.  As for $\nu=1$, the $K=0$ and
$K \to \infty$ limits of the charged excitation energy
can\cite{wu} be calculated analytically when finite-thickess
corrections are neglected.  Wu and Sondhi\cite{wu}
have recently pointed out that for $n \ge 1$,
$U_{K \to \infty} >  U_{K=0}$ so that Hartree-Fock quasiparticles
have lower energy than large spin-texture quasiparticles even
when the Zeeman energy is not included.
Recent transport experiments~\cite{schmeller}
seem to imply that for $\nu = 3$,
the lowest energy charged excitations have $K=0$,
in contrast with the $n=0$ case and in agreement with theory.
In this section we examine the influence of the finite
width of the electron layer on the
energetics of the quasiparticles.

The competition between Hartree-Fock and charged spin-texture
quasiparticles can be understood in terms of the
analytically known expressions for
the Hartree-Fock exchange energy per electron for a full Landau level
($\epsilon_{0}$ in Eq.(~\ref{eq:epsilon_{0}})) and for the
spin-stiffness\cite{sondhi,moon}
\begin{equation}
\rho_{s}=\frac {1}{16\pi} \int \frac {d^{2}q}{(2\pi)^{2}}
q^{2} v_{n}(q) e^{-q^{2}/2}.
\label{eq:rhos}
\end{equation}
Since $U_{K=0} =  - 2 \epsilon_{0}$ and
$U_{K \to \infty} = -\epsilon_{0} + 4 \pi \rho_{s}$ the
difference is
\begin{equation}
\Delta U \equiv U(K=0) - U(K=\infty)
 =  -\epsilon_{0} - 4 \pi \rho_{s}
 =  \frac{1}{4}  \int \frac {d^{2}q}{(2\pi)^{2}}
[2 - q^{2} ]v_{n}(q) e^{-q^{2}/2}.
\label{eq:endiff}
\end{equation}
For the Coulomb interaction model\cite{wu}
$\Delta U = (-1/16) (\pi/2)^{1/2} (e^{2}/\ell) $
in the $n=1$ ($\nu =3$) Landau level and has a larger negative
value for the $n=2$ ($\nu =5$) case.
The $[L_{1}(q^{2})]^{2} $ form factor for electrons in the
$n=1$ Landau level strengthens the
effective interaction at large $q$ which has
more importance in $\rho_{s}$ than in $\epsilon_{0}$
because of the $q^{2}$ factor in its integrand.  When
finite thickness form factors are also included in the
effective interaction, the importance of large $q$ in
these integrals will diminish.  For sufficiently wide quantum
wells it is clear that the sign of $\Delta U$ will be
positive and charge spin textures will
also occur in the $n=1$ Landau level.
The objective of the calculations reported on below is
to quantify the stability properties of skyrmions in $n=1$
quantum Hall ferromagnets with finite width electron layers.
We find that skyrmions in the first Landau level
can be stabilized by finite thickness corrections,
although for realistic widths the maximum $\tilde{g}$ for which they
are the lowest energy quasiparticles is roughly
an order of magnitude smaller than for the $n=0$ case.
Thus the observation of skyrmions in the first Landau level would require
specialized samples or experimental
techniques to access this very low Zeeman energy limit.

Figure~\ref{fig:five} illustrates the energy $\epsilon^{-}_{K}$
for both the $K=0$ quasiparticle (dotted line) and the
optimal $K>0$ skyrmion (solid line) as a function of
$\tilde{g}$ for a typical experimental sample width,
$W=0.45\ell$.  The skyrmion excitation is
lower in energy for $\tilde{g}<2.4\times 10^{-3} e^{2}/\epsilon \ell$.
The results in Figure~\ref{fig:five} suggest that for filling factors
close to, but slightly away from, $\nu=3$, a strong first-order
phase transition should take place as $\tilde{g}$ is increased from zero,
in which the spin polarization changes very abruptly.
That this transition should involve a very large number
of flipped spins can be seen by noting that the optimal
values of $K$ of the skyrmions for small values $\tilde{g}$
is extremely large, as illustrated in Fig.~\ref{fig:six}.
Unlike the $n=0$ case, skyrmion solutions of the Hartree-Fock
equations become unstable at a finite value of $K$ ($ \approx 9$ for
$W = 0.45 \ell$)  at
which $ \tilde{g} = -dU(K)/dK$ reaches its maximum value
($\approx 0.003 e^{2}/\epsilon \ell$ for $W=0.45 \ell$).
For $\tilde{g}$ larger than this value there are no $K \ne 0$ solutions
of the Hartree-Fock equations.

We estimate the number of flipped spins in the skyrmion state for
$W=0.45\ell$ at the critical value of $\tilde{g}$
to be approximately $K=14$, using our Hartree-Fock
approach with a system size of $M_{\rm max}=480$.  It should
be noted that at finite skyrmion concentrations, especially
for such large skyrmions, a
calculation\cite{skyrmlat} including interactions
among the skyrmions would give a more reliable estimate
for the jump in the spin polarization at the transition.
We expect that such interactions would most likely reduce somewhat the
magnitude of this jump. (Related finite size studies have been
carried out by Jain and Wu \cite{wu}.)

In general, the stability of $K>0$ skyrmions is most easily
assessed by considering the internal energy $U_{K}$.
As in the previous section, $U_{K}$ may be computed numerically
by finding $\epsilon^{-}_{K}$ using the Hartree-Fock
method, and then subtracting off the Zeeman contribution
to the energy, as in Eq.~\ref{eq:epsilonkm}.  The results of
such a calculation are illustrated in Fig.~\ref{fig:seven}
for several values of the layer thickness $W$. These results
confirm that skyrmions are the lowest energy charged excitations
for large enough layer thickness, but become higher in energy than
the $K=0$ quasiparticle as the system approaches the
two-dimensional limit. For n=1, $U_{K}$ may have local minima
both at $K=0$ and at $K \rightarrow \infty$. For this reason it is possible for
the Hartree-Fock equations to have two separate
solutions for the same values of $\tilde{g}$,
as shown in Fig.~\ref{fig:five}. (In the
$NL\sigma$ model calculation \cite{wu}, the energy of the skyrmion branch
{\em increases} from that of the $\tilde g=0$ solution in precisely the
same fashion as in the LLL case for there is no distinction
between the direct Coulomb interaction in the two Landau levels
for skyrmions of divergent size. From the independent calculation of the
lower energy polarized quasiparticle, it follows that the skyrmion energy
must be non-monotonic with size and, 
in the absence of intervening additional extrema, must
have minima at both microscopic and infinite size.)
It is interesting to note that, unlike the case of skyrmions in the
lowest Landau level, the curvature of $U_{K}$ changes with increasing $K$
(cf. Fig.~\ref{fig:three}).  As mentioned in the discussion following
Eq.~\ref{eq:selfcons}, we do not expect to find stable skyrmion solutions
for the convex regions of the graphs in Fig.~\ref{fig:seven}.
Indeed, in order to obtain values for $U_{K}$ in the small
$K$ limit, it is necessary to add a fictitious term
to the Hamiltonian of
the form $H_{\alpha}=\alpha(\hat{S}_z-S_z^0)^{2}$, where
$\hat{S}_z$ is the operator for the
total spin angular momentum of the
state.  Since this term couples only to the total spin
of the system, it favors states with  $\langle \hat{S}_z \rangle
\approx S_z^0$, but does not affect the optimal shape of
the spin texture within the subspace of states with this
expectation value for the spin.  Thus by varying both
$\alpha$ and $S_z^0$ we can obtain states with small
values of $K$, and their intrinsic energy may be
obtained by subtracting the expectation value $\langle
H_{\alpha} \rangle$ from the expectation
value of the perturbed total Hamiltonian.
We note that this procedure produces internal energies
that join smoothly onto those obtained for larger
values of $K$ where the skyrmion states are stable
for large enough layer thickness.

Our results for this section are summarized in Fig.~\ref{fig:eight},
by a ``phase diagram'' in $\tilde{g}$ and $W$
which shows where the lowest energy charged excitations are
$K=0$ quasiparticles and where they are
$K \ne 0 $ skyrmion charged spin-textures.
Even for the most favorable
layer thickness, $W \approx 1.2\ell$, the maximum value
of $\tilde{g}$ for which skyrmions are stable is outside the
accessible range in many typical experimental systems.
Nevertheless, we emphasize that
if a system with very small effective values of
$\tilde{g}$ could be fabricated, for values of
$\nu$ close to (but not exactly equal to) 3
this line will separate states of maximal polarization ($K=0$)
from states with highly degraded polarizations due to the
presence of skyrmion spin textures.

\section{Summary}

In this work we have studied the charged skyrmion
excitations of quantum Hall ferromagnets
within the Hartree-Fock approximation.
Exact relationships between the energies and spin quantum
numbers of quasihole and quasiparticle
excitations energies near $\nu=1$ were derived using
particle-hole symmetry.  An internal energy function $U(K)$
was introduced to describe the relative stability competition between
skyrmions states with different numbers of reversed spins, $K$.
Locally stable skyrmion states with $K$ flipped spins can occur
for any Zeeman coupling strength only if $d^2 U(K)/ d K^2 > 0$.
In the presence of Zeeman coupling, a stable skyrmion states
must satisfy $d U(K)/ dK = - g^{*} \mu_B B$.
Results for the dependence of the $K$ value of
the lowest energy skyrmion excitations on
Zeeman coupling strengths were presented both with
and without finite thickness corrections.  Finite
thickness corrections were shown to stabilize
skyrmions in the $n=1$ Landau level,
where for zero thickness $K=0$ quasiparticles are lowest in energy
at any Zeeman coupling strength.   We have proposed that a first
order phase transition between states with
relatively large skyrmions and states with $K=0$ quasiparticles
will occur at $\nu =3$ at a critical value of the
Zeeman coupling $\tilde{g}$ and be
accompanied by a jump in the spin-magnetization of the electron system.

\acknowledgements

This work was supported in part by  NATO Collaborative Research Grant No.
930684, by the National Science Foundation under grants DMR-9416906,
DMR-9503814 and DMR-9632690, CICyT of Spain under contract MAT 940982
and by the Swedish Natural Science Research Council. Additional support
was provided by the A.P. Sloan Foundation (HAF \& SLS) and the Research
Corporation (HAF).  Helpful conversations with Steven Girvin, Steven
Kivelson, Luis Martin-Moreno, Kyungsun Moon,
Juan-Jose Palacios, Carlos Tejedor, Edward Rezayi and Kun Yang are
gratefully acknowledged.

\bigskip
\noindent {\it Note added:} After the completion of this work, we
became aware of a study of finite thickness effects on skyrmions\cite{cooper}
for electrons on a sphere using an approximate Gaussian form for the
confined state $\chi(z)$ of the electrons.   Where there is
overlap, the results presented there
agree reasonably well with our own.

\appendix
\section{Midgap States and the Infinite Skyrmion}

The most intuitive account of the physics of the skyrmions relies
upon the spin Berry phases generated by adiabatic motion in a
textured (effective) magnetic field arising from the magnetic
exchange among the electrons themselves. Geometrical considerations
show that the Berry phases due to a unit topological charge precisely
mimic those of an additional quantum of {\em orbital} flux. The
incompressibility of the quantum Hall fluid then implies that an
additional electron (or fraction thereof for fractional quantum
Hall states) must
then be present in the region of the texture which is therefore a
(dressed) quasielectron.

A more detailed description of this is to note that ``up'' and ``down''
spins, i.e. the spins that respectively point parallel and anti-parallel
to the {\em local} direction of the field in the texture, see oppositely
signed fluxes. Consequently while the ``up'' spin (effective) Landau level
gains one state the ``down'' Landau level loses one, i.e. the presence of
the skyrmion causes the transfer of one state beween the upper (empty) and
lower (filled) bands at $\nu=1$ thus allowing the extra electron to be
accomodated in a low energy state, albeit at the energetic cost of texturing
the spins. This is strongly reminiscent of midgap states in one-dimensional
systems, i.e. polyacetylene \cite{js}, where a background soliton causes states
to appear in a gap and again there is the cost of creating the soliton which
has to be overcome.

Since this description is very much of the one-electron variety, it is
instructive to see how it works out in the the Hartree-Fock treatment.
The situation is clearest for the ``infinite'' skyrmion where the texture
is arbitrarily slowly varying. It is very useful to regularize this
limit by putting the system on a sphere where it takes the form of a
purely radial texture, ${\bf n}(\theta, \phi)={\bf \hat{r}}$. Again we
may deduce the Hartree-Fock description by noting that the classical texture
is invariant under the action of ${\bf J = L + S}$. In the LLL
approximation this uniquely picks the wavefunction and fixes the
eigenoperators of the Hartree-Fock Hamiltonian. For a sphere with
$2 q$ flux quanta threading the surface, the latter are:
\begin{eqnarray}
c^{\dagger}_{j+} &=& u_j a^{\dagger}_{j-1/2} + v_j b^{\dagger}_{j+1/2},
\ \ \ |j| \le q+1/2
\nonumber \\
c^{\dagger}_{j-} &=& v_j a^{\dagger}_{j-1/2} - u_j b^{\dagger}_{j+1/2},
\ \ \ |j| \le q-1/2
\end{eqnarray}
where
\begin{eqnarray}
u_j = \sqrt{q+1/2+j \over 2q+1}  \ \ {\rm and} \ \
v_j= \sqrt{q+1/2-j \over 2q+1} \ .
\end{eqnarray}
In this basis the Hamiltonian and wavefunction are,
\begin{eqnarray}
H_{HF} &=& \sum_{|j| \le q-1/2} \epsilon_{-} c^{\dagger}_{j-} c_{j-}
          -\sum_{|j| \le q+1/2} \epsilon_{+} c^{\dagger}_{j+} c_{j+}
\nonumber \\
|\Psi\rangle_{HF} &=& \prod_{|j| \le q+1/2} 
c^{\dagger}_{j+} c_{j+} |0\rangle, \ 
\end{eqnarray}
where $|0\rangle$ is the vacuum state.
The energies $\epsilon_\pm$ depend upon the details of the
interaction and not just on the symmetries of the state. (In the
limit of infinite system size these become degenerate with the
corresponding up and down spin
eigenvalues at (exactly) $\nu=1$ as the skyrmion
becomes locally indistinguishable from the ferromagnetic state.)
The form of $H_{HF}$ clearly shows that transfer of one state across
the gap. Evidently, all we have used in this is that the states of
$H_{HF}$ must be classified only by their eigenvalues under $\bf J$
and hence must occur in the multiplets $j+1/2$ and $j-1/2$. This
will therefore also survive the perturbative inclusion of Landau level
mixing.

For finite skyrmions, the picture is (unfortunately!) less elegant. The
excess charge is now localized in a finite region and hence affects the
Hartree-Fock eigenvalues in the core of the skyrmion. This can be seen
in from Fig.~\ref{fig:nine}, where we have plotted the eigenvalues,
now on the plane,
for skyrmions at two different values of $\tilde g/ (e^2/\epsilon \ell)$.
While the general scenario of a transferred state between two sets of levels
still holds, the individual
sets themselves become degenerate only in the large
skyrmion limit.

Finally, we note that putting the infinite skyrmion on the sphere allows
us to compute its energy very accurately and thus circumvent the problems
at small $\tilde g$ noted in the text. The only trick here is that we
need to subtract the Hartree self-interaction of the skyrmion charge,
$e^2/(2 \sqrt{q})$ ($\sqrt{q}$ is the radius of the sphere), which
is substantial for small system sizes. This greatly improves convergence
and the known infinite system result is recovered to within a few percent
already at system sizes with $q \approx 10$.

\begin{figure}
\caption{Charge and spin densities for quasihole skyrmions in
the lowest Landau level for a system of zero thickness, evaluated
for different values of $\tilde{g}$.  Maximum angular momentum
state kept in these computations was $M_{\rm max}=120$ (see text).}
\label{fig:density}
\end{figure}

\begin{figure}
\caption{Minimum hole creation energy (solid line)  and transport
activation energy (dashed line)  in a $\nu =1$ quantum Hall ferromagnet as
a function of $\tilde{g}$.
All energies are in units of $(e^{2}/\epsilon \ell)$ and
results are shown for both strictly 2D and finite-thickness models.
Maximum angular momentum state in these calculations was
$M_{\rm max}=120$.}
\label{fig:one}
\end{figure}

\begin{figure}
\caption{Number of reversed spins per hole in a
$\nu =1$ quantum Hall ferromagnet as a function of
$\tilde{g} / (e^{2}/\epsilon \ell)$. Results are shown for both strictly 2D
and finite-thickness models. Note that for self-consistent solutions
of the Hartree-Fock equations $\tilde g = - d U(K)/ dK$.
Maximum angular momentum state in these calculations was
$M_{\rm max}=120$.}
\label{fig:two}
\end{figure}

\begin{figure}
\caption{Internal energy of a skyrmion hole $U$ as a function of the
number of reversed spins $K$.   Results are shown for both strictly 2D and
finite-thickness models.
Maximum angular momentum state in these calculations was
$M_{\rm max}=120$.}
\label{fig:three}
\end{figure}

\begin{figure}
\caption{Number of reversed spins in a skyrmion hole at weak
Zeeman coupling as a function of finite system size.}
\label{fig:four}
\end{figure}

\begin{figure}
\caption{Energy $\epsilon^{-}_{K}$ of the $K=0$ quasiparticle
(dotted line)
and the HF $K>0$ skyrmion (solid line)
as a function of $\tilde{g}$ for $\nu =3$ ($n=1$)
and a 2DES with thickness parameter $W=0.45\ell$.
The system size used for this calculation was $M_{\rm max}=480$.
All energies are in units of
in units of $e^{2}/\epsilon\ell$.}
\label{fig:five}
\end{figure}

\begin{figure}
\caption{
Number of flipped spins $K$ of the HF $K>0$ skyrmion
as a function of $\tilde{g}$,
for $\nu =3$ ($n=1$) and a 2DES with thickness $W=0.45\ell$.
Note that for self-consistent solutions
of the Hartree-Fock equations $\tilde g = - d U(K)/ dK$.
The system size used for this calculation was $M_{\rm max}=480$.
All energies are in units of $e^2/\epsilon\ell$.}  Vertical
dashed line indicates the value of $\tilde g$ above which
the skyrmion is unstable against the untextured quasihole.
\label{fig:six}
\end{figure}

\begin{figure}
\caption{
Internal energy $U_{K}$ vs. $K$ for skyrmions in
the $n=1$ Landau level systems of varying thicknesses $W$.  Skyrmions
are the lowest energy quasiparticles for sufficiently weak
Zeeman coupling if $U_{K\rightarrow \infty}<U_{K=0}$.
We find two distinct solutions to the Hartree Fock equations
for $W=0.1\ell$ in the vicinity $K=13$, leading to a cusp
in $U_{K}$; for larger values of $W$ only one solution
is found and the cusp becomes a smooth minimum.  Where
$d^2 U(K) / d K^2 < 0$ solutions to the Hartree-Fock equations
can be found only by adding unphysical terms to the Hamiltonian as
discussed in the text.  The system size used
for these calculations was $M_{\rm max}=480$. Energies are in
units of $e^2/\epsilon\ell$.}
\label{fig:seven}
\end{figure}

\begin{figure}
\caption{
Phase diagram for the dependence of the
globally stable Hartree-Fock quasiparticle at $\nu =3$ ($n=1$)
on system parameters $\tilde{g}$
and layer width $W$.  ($\tilde g$ is in units of $e^2/\epsilon\ell$ and
$W$ is in units of $\ell$.)  The phase boundary
separates parameter values for which the $K=0$ quasiparticle is
lowest in energy from parameter values for which
$K>0$ skyrmions are lowest in energy.
Systems in which these parameters can be adjusted
so as to cross the phase boundary should exhibit
a jump in the magnetization.  These calculations were performed
with $M_{\rm max}=480$.}
\label{fig:eight}
\end{figure}

\begin{figure}
\caption{
The Hartree-Fock eigenvalues for skyrmions in the LLL at
$\tilde g/(e^2/\epsilon \ell)=0.1$ (stars) and $0.010$ (circles).
The former is the $K=0$ quasiparticle and the latter has
$K \approx 6$. In both cases there is an extra state in the
lower branch. As there are 301 orbitals in the system
$M_{\rm max}=300$, this means that
the upper branches have 300 and the lower (occupied) branches have
302 states. Note that the gap is smaller near the origin which
is where the extra charge resides and near the boundary due to
loss of exchange. The shift in the eigenvalues near the origin is
due to the Hartree repulsion of the extra charge. Evidently,
the eigenvalues are much more uniform for the skyrmion
but the overall reduction of the gap in its case is a finite
size effect.
}
\label{fig:nine}
\end{figure}

\begin{references}
\bibitem{fn1} It is an interesting feature that the pure sigma model is
apparently realized {\em exactly}, by the hard-core pseudopotential model with
$V(0)\ne 0$. This conclusion follows from the observation
\cite{ednumerical,hcmskyrm} that the entire family of ``skyrmion''
quasiparticle states with $L=S$
are degenerate in this case; in the sigma model language this implies scale
invariance for the action. It was also noted in \onlinecite{hf} that the
hard-core skyrmions are described exactly by the Moon et al \cite{moon}
wavefunctions which reproduce the classical textures for large sizes. This
circle is closed by the further observation in that paper that the
hard core skyrmions can be obtained as eigenfunctions of the quadratic
hamiltonian $H=- P {\bf n}(x)\cdot {\bf S}(x)P$, where P is the LL projection
operator, {\bf n} the classical texture and ${\bf S}$ is the second quantized
spin operator.
\bibitem{comhf} For quasielectrons or anti-skyrmions, the spin and charge
densities are identical (except for the sign of the charge
density) to those shown in Fig.~\ref{fig:density} due to
particle-hole symmetry.  Note that a similar illustration
appears as Fig. 2 in Ref.~\onlinecite{hf}.  However, that
figure contains a numerical error in the computation of
the densities for the quasiparticle, and is mislabeled
for the quasihole skyrmion densities.
\bibitem{comhf2} The definition of $\tilde{g}$ here differs
slightly from that of Ref.~\onlinecite{hf}.  In that work $\tilde{g}$
was the coefficient of the Zeeman coupling term of the
Hamiltonian in units of $e^2/\epsilon\ell$.
Here we take $\tilde{g}$ to be the energy of
the Zeeman splitting for non-interacting electrons.
\end{references}
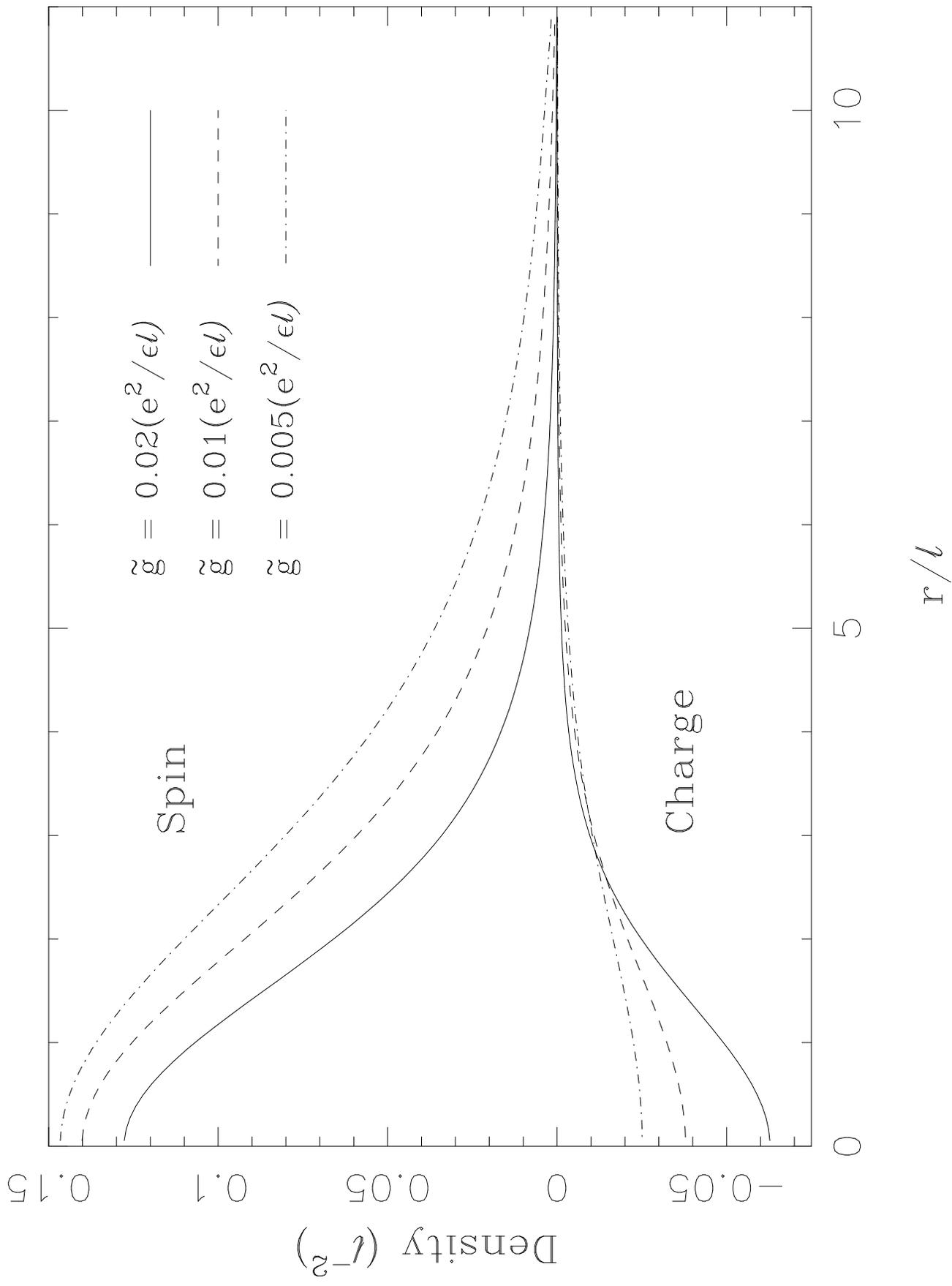
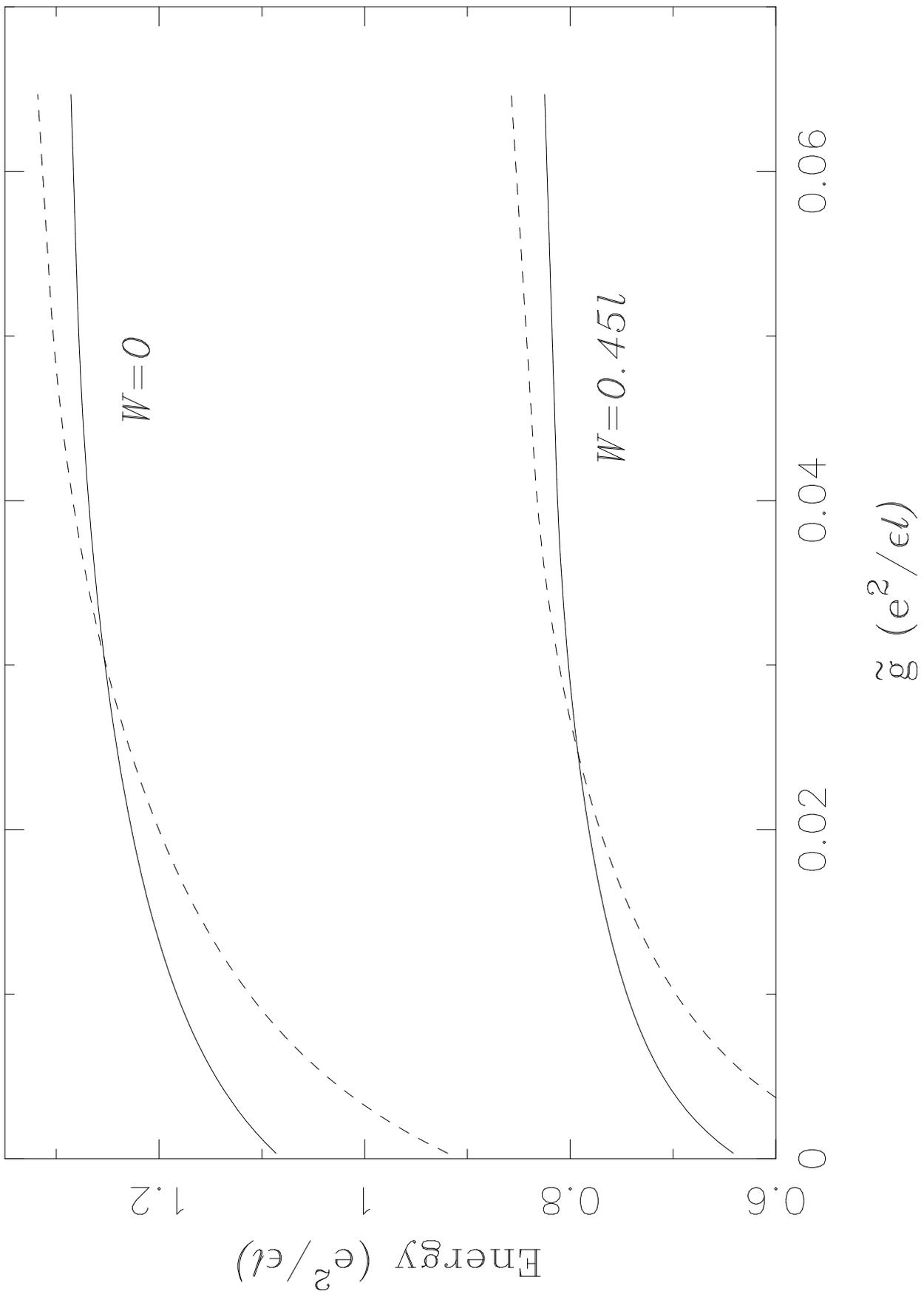
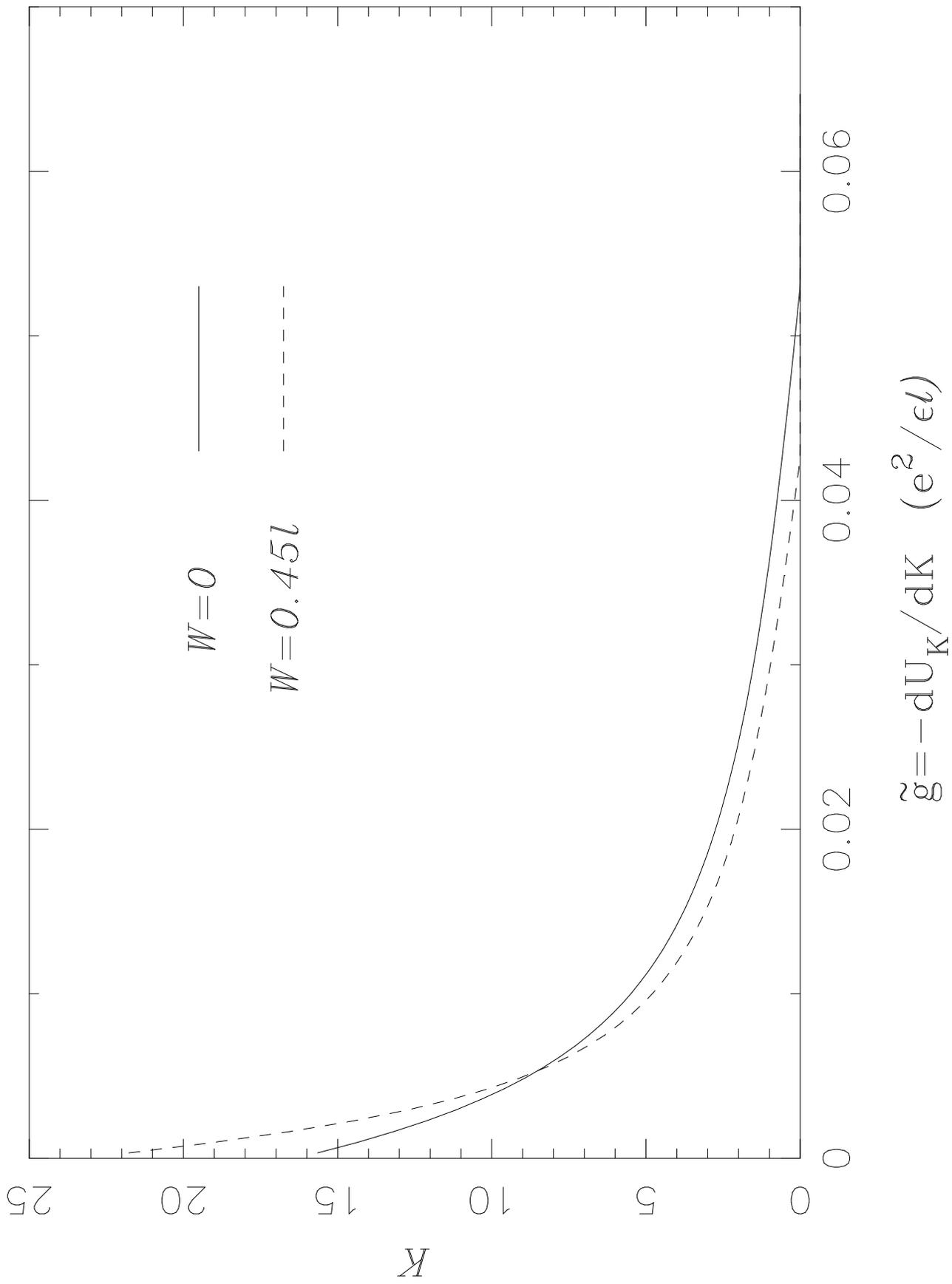
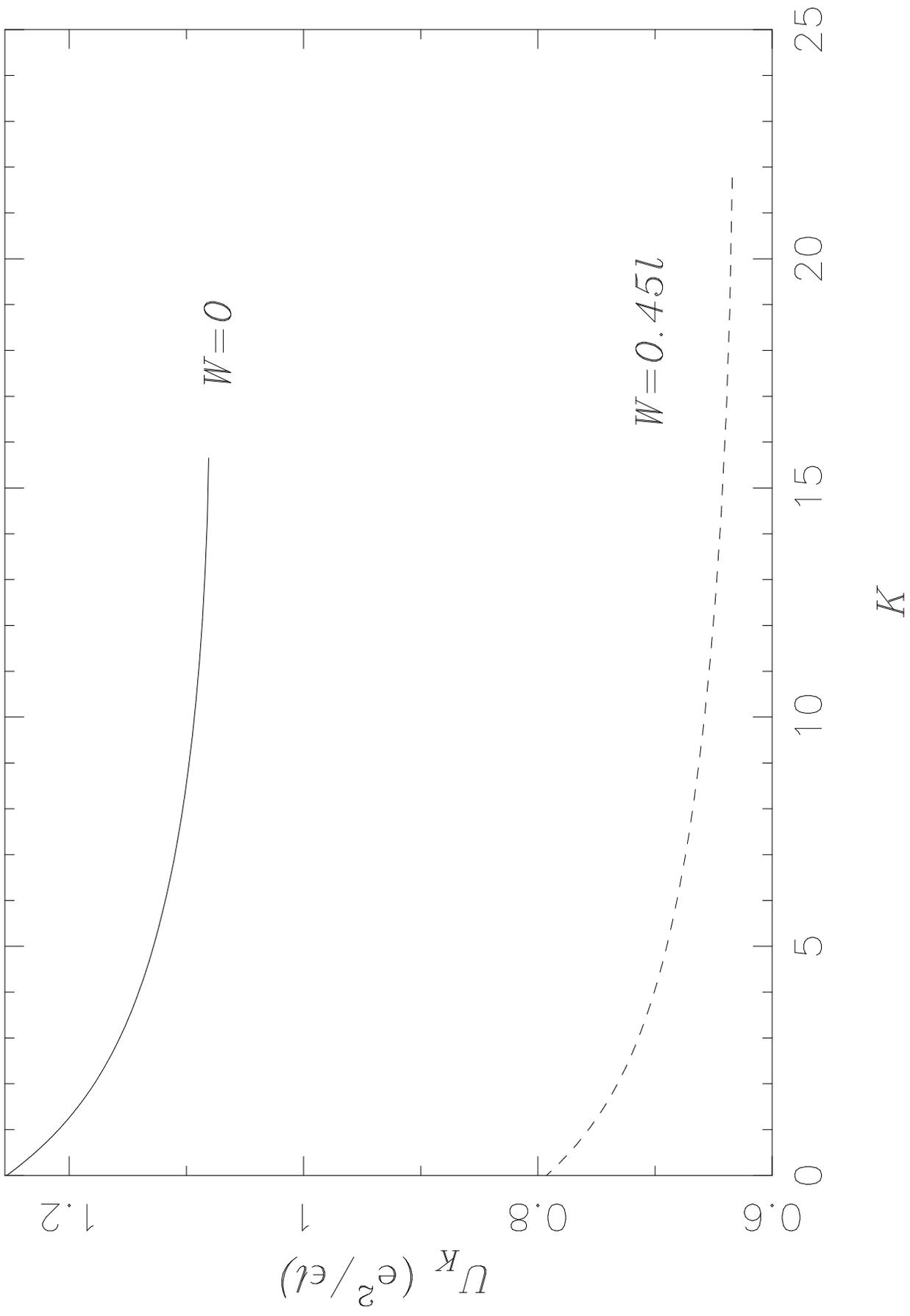
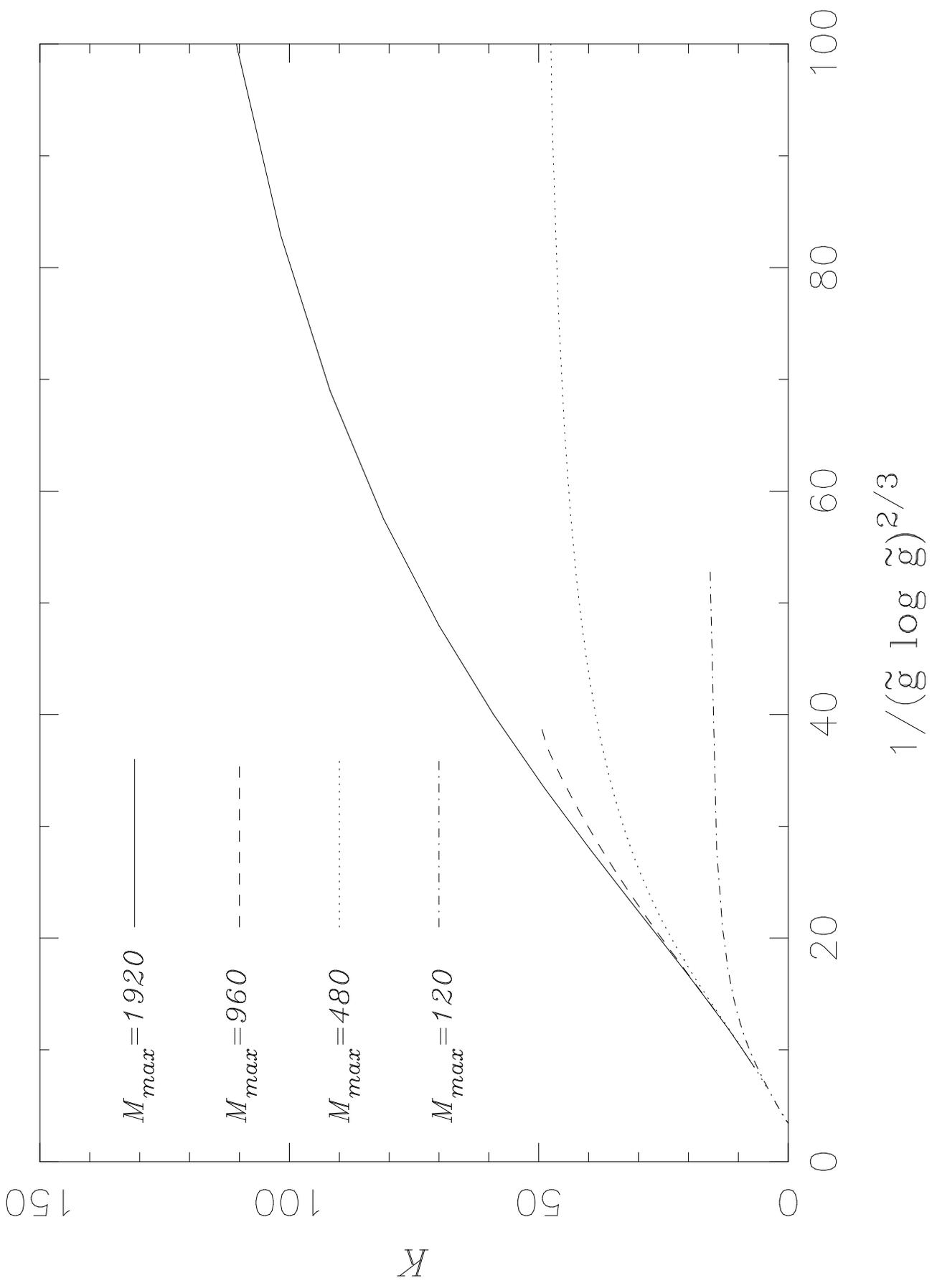
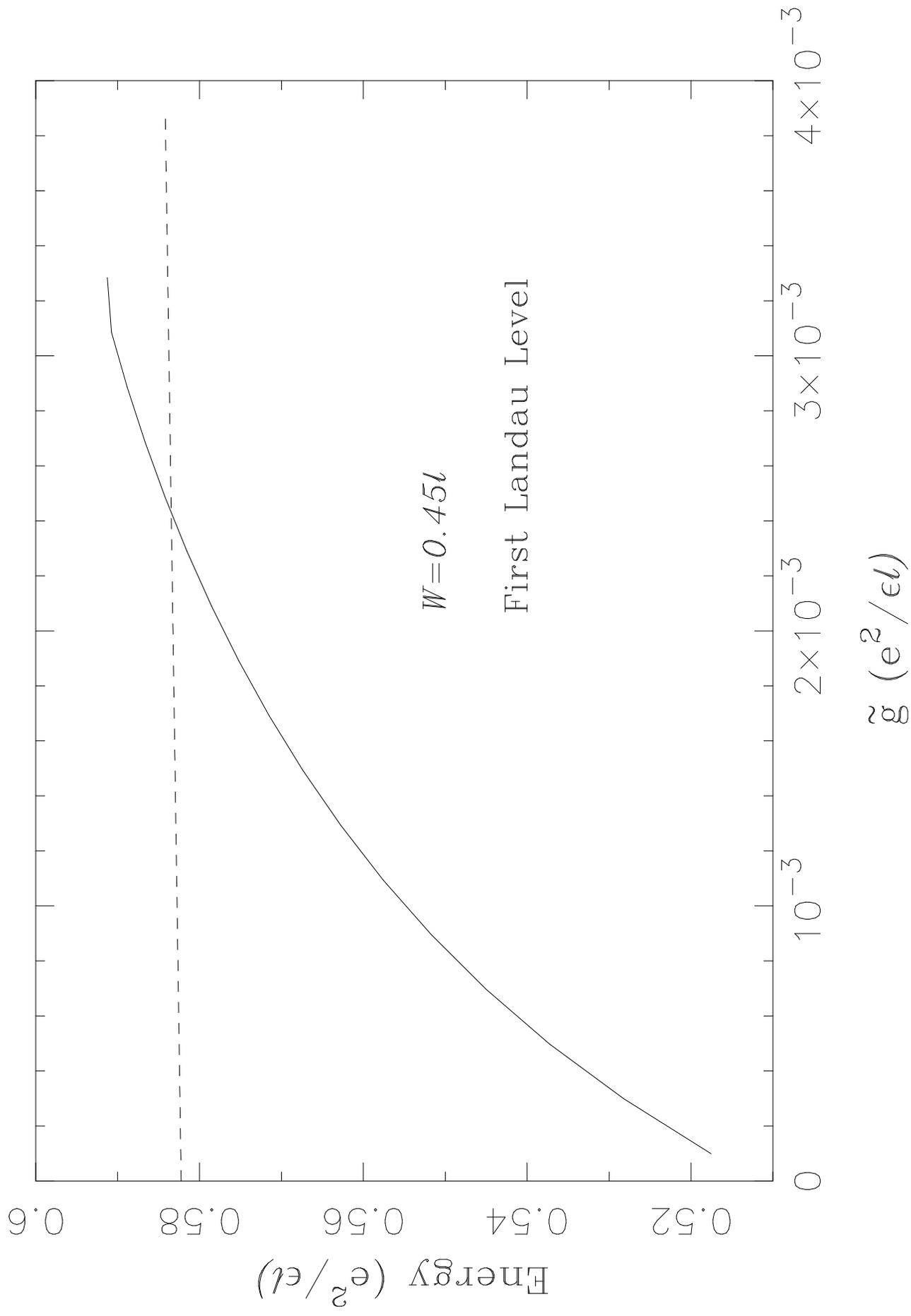
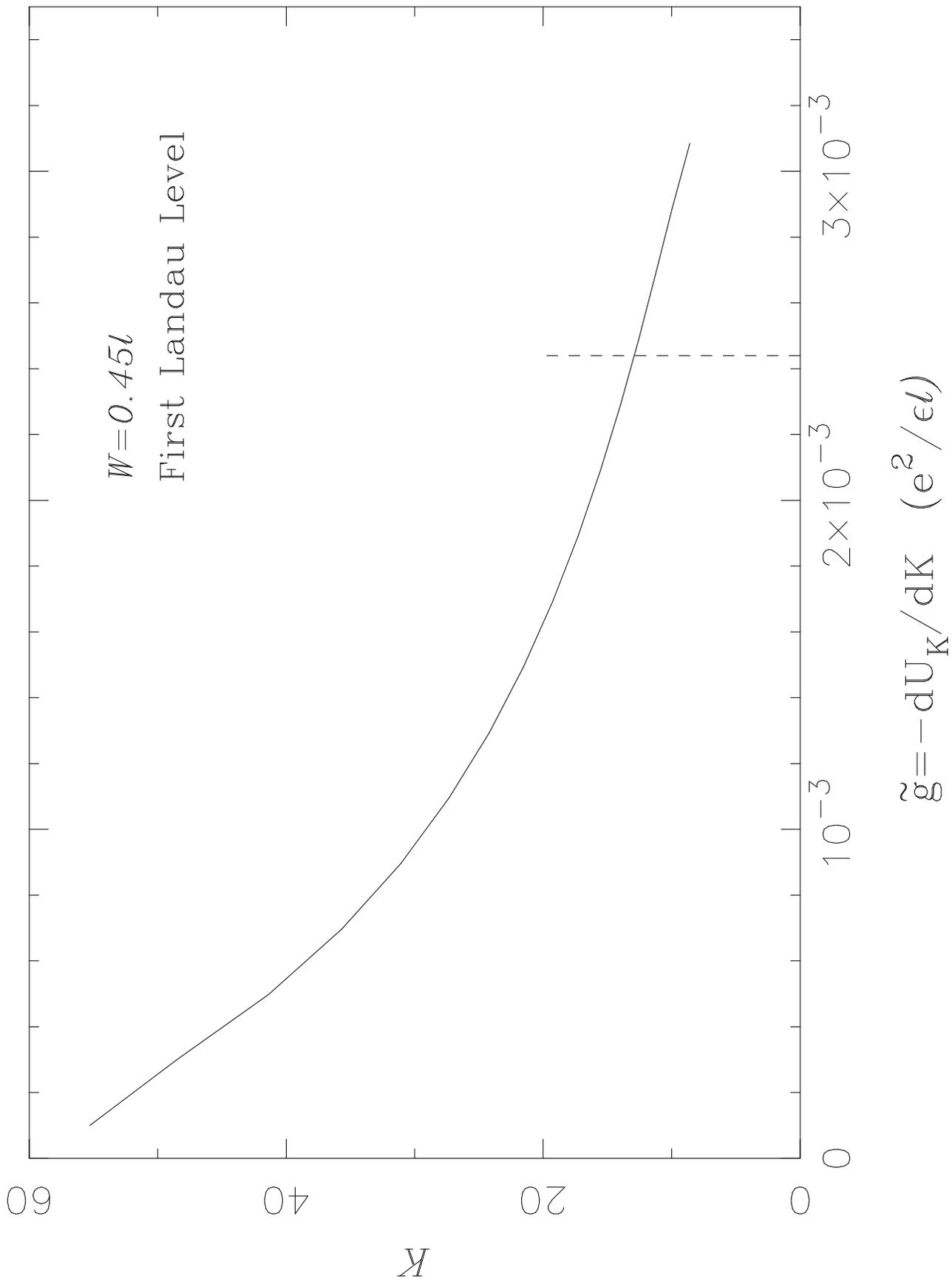
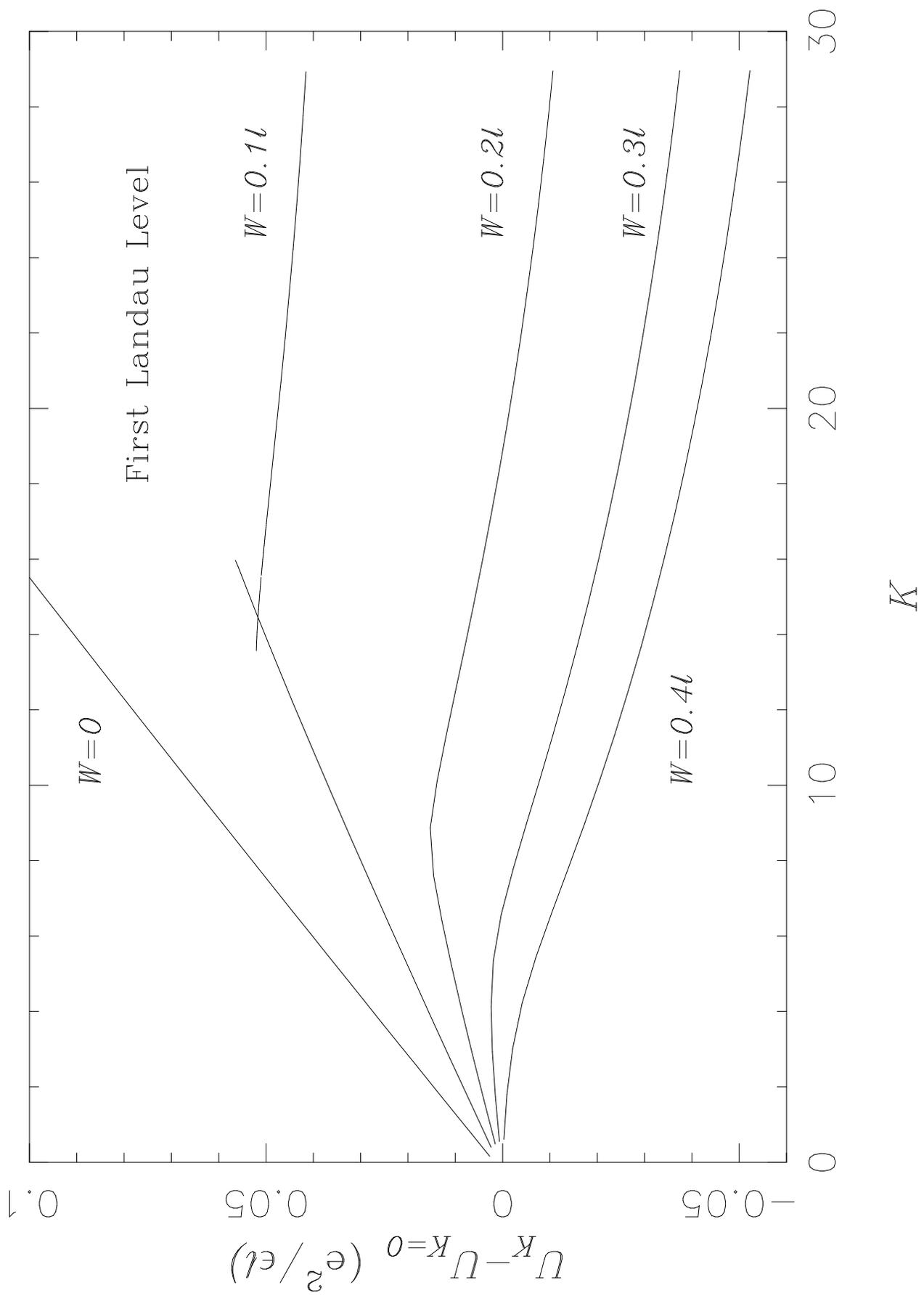
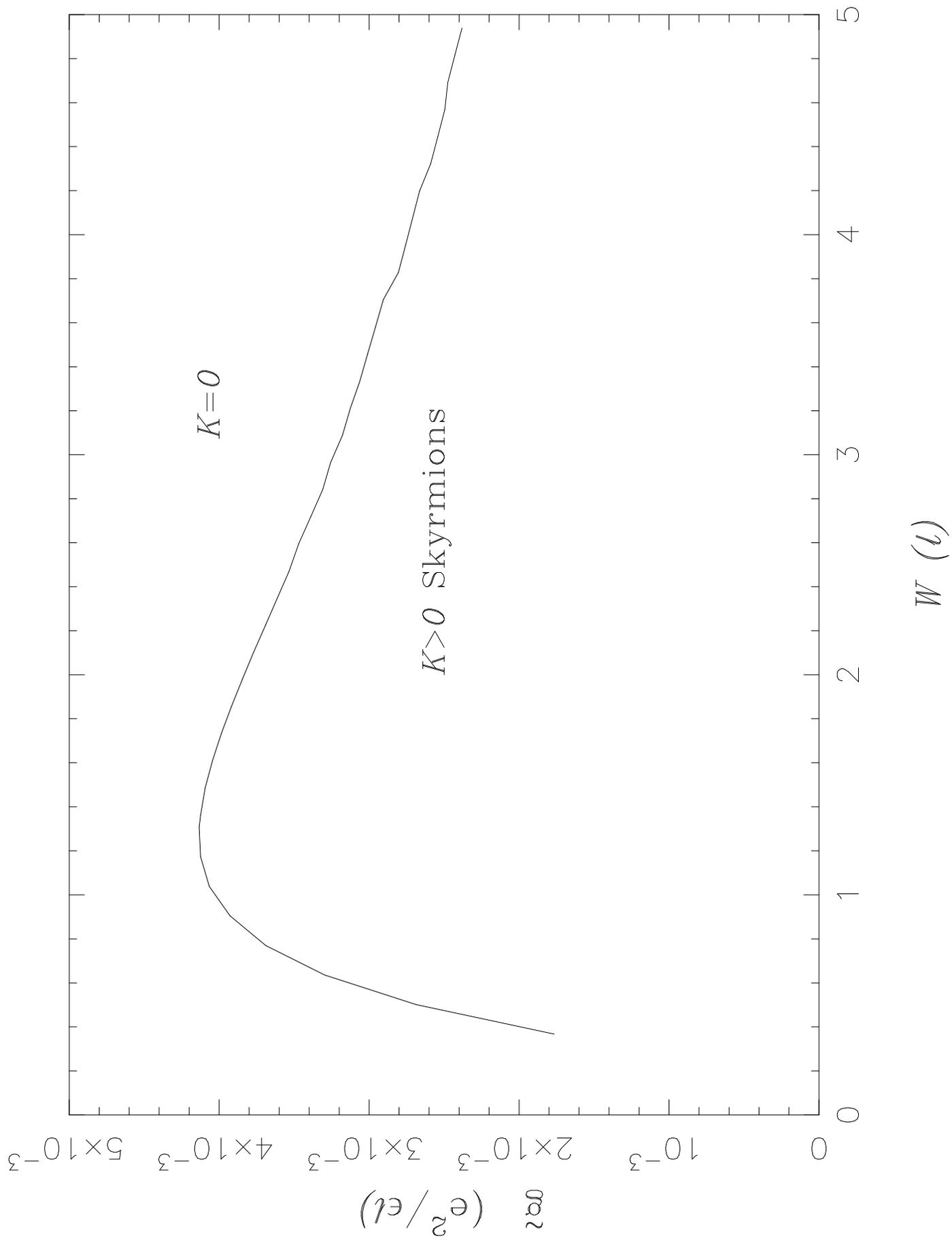
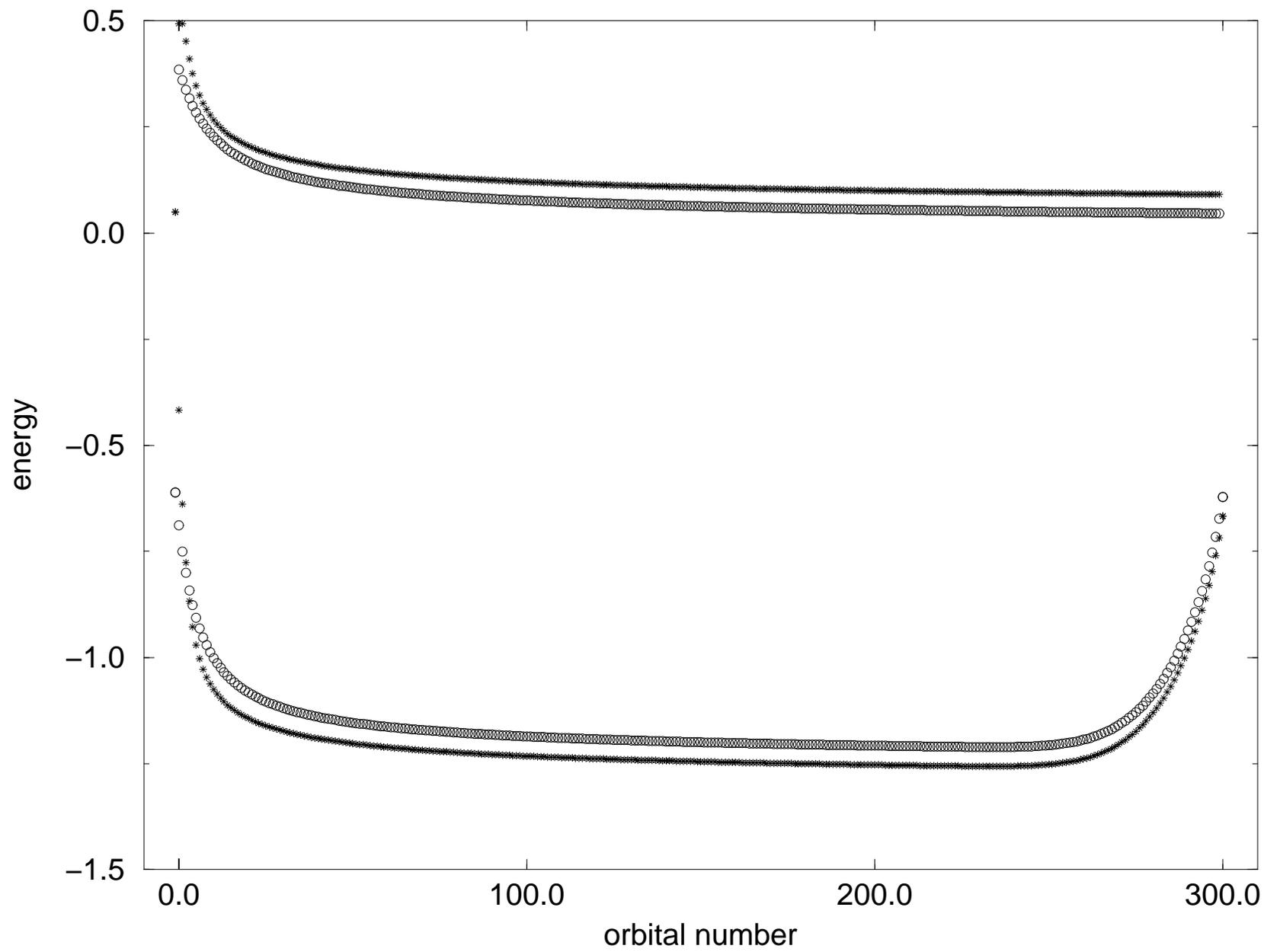
\end{document}